\documentstyle[aps,prl,epsfig,fancyheadings]{revtex}

\begin{document}

\twocolumn[
\hsize\textwidth\columnwidth\hsize\csname@twocolumnfalse\endcsname

\draft

\title{Linear Response Calculations of Lattice Dynamics in Strongly Correlated
Systems}
\author{S. Y. Savrasov}
\address{Department of Physics, New Jersey Institute of Technlogy, Newark, NJ 07102}
\author{G. Kotliar}
\address{Department of Physics and Astronomy, Rutgers University, Piscataway, NJ 08854}
\date{April 2002}
\maketitle

\begin{abstract}
We introduce a new linear response method to study  the lattice
dynamics of materials with strong correlations. It is based on a
combination of dynamical mean field theory of strongly correlated
electrons and the local density functional theory of electronic
structure of solids. We apply the method to study the  phonon
dispersions of a prototype Mott insulator NiO. Our results show
overall much better agreement with experiment than the
corresponding local density predictions.
\end{abstract}

] 
\narrowtext

Computational studies of lattice dynamics and structural stability in
strongly correlated electronic systems is a challenging theoretical problem.
In the past density functional theory (DFT) in its local density or
generalized gradient approximations (LDA or GGA) \cite{DFTbook} has
delivered the full lattice dynamical\ information and electron-phonon
related properties of a variety of simple metals, transition metals, as well
as semiconductors with exceptional \ accuracy\cite{BaroniRMP}. This is
mainly due to the introduction \ of a linear response approach \cite
{BaroniPRL,Zein}. This method overcame the problems of traditional
techniques based on static susceptibility calculations which generally fail
to reproduce lattice dynamical properties of real materials due to
difficulties connected with the summations in high--energy states and the
inversion of very large dielectric matrix \cite{Doren}.

Despite these impressive successes, there is by now clear evidence that the
present methodology fails when applied to strongly correlated materials. For
example, the local density predictions for such properties as bulk modulus
and elastic constants in metallic Plutonium are approximately one order of
magnitude off from experiment\cite{Bouchet}; the phonon spectrum of Mott
Insulators such as MnO is not predicted correctly by LDA \cite{Massida}.

In this work we describe a new linear response method to study the lattice
dynamics of correlated materials. It is based on the dynamical mean field
theory (DMFT) \cite{DMFTreview}, a many-body technique developed to study
systems with strong on-site Coulomb repulsion. Recent progress in merging 
\cite{Anisimov} this many-body description with the realistic LDA based
electronic structure calculations has already led to\ new insights in long
standing problems, such as, the temperature dependence of the magnetic
properties of Fe and Ni \cite{Licht}, the volume collapse transition in Ce 
\cite{CeVolume}, and Pu \cite{Nature,Functional}. We generalize this
LDA+DMFT method to carry out linear response calculations by finding
self--consistent changes in both charge densities and local Green functions
induced by atomic displacements.

The LDA+DMFT\ approach computes the total energy, the charge density and the
local spectral function of the correlated electrons simultaneously. The
latter spectra are known to be quite different for a strongly correlated
system such, e.g., as heavy fermion metal, from the Kohn Sham spectra of the
local density functional theory due to the appearance of strongly
renormalized quasiparticle features at the Fermi level and lower and upper
Hubbard features or satellites at higher energies\cite{DMFTreview}. The
LDA+DMFT\ technique can therefore provide a link between the photoemission
spectra and the lattice dynamics of correlated materials, which is an
interesting open problem.

As a test we consider the prototype Mott insulator NiO. In this material at
low temperatures, the static Hartree--Fock limit of our formulation which is
formally equivalent to the LDA+U method \cite{LDA+U} can be used. Results
for the phonon dispersion and the static dielectric properties of this
material are found in much better agreement with the experiment than the
results of corresponding LDA based calculations. We discuss how correlations
affect the calculated dielectric constants and the Born effective charges
using the linear--response method.

Our approach considers both the charge density $\rho $ and the local Green
function $\hat{G}(\omega )$ as parameters of a spectral density functional 
\cite{Functional}. To find its extremum, a set of Dyson equations is solved
self--consistently: 
\begin{equation}
(-\nabla ^{2}+V_{eff}+\hat{\Sigma}(\omega )-\hat{\Sigma}_{dc}-\epsilon _{%
{\bf k}j\omega })\psi _{{\bf k}j\omega }^{r}=0  \label{Dyson}
\end{equation}
where $V_{eff}$ is the effective potential of the local density functional
and $\hat{\Sigma}(\omega )$ is the local self--energy operator. Since the
local density approximation contains an average correlation energy, a double
counting term $\hat{\Sigma}_{dc}$ appears in the equation (\ref{Dyson}).
Both $V_{eff}$ and $\hat{\Sigma}(\omega )$ are functionals of the density
and the local Green function which can be found using the formula:

\begin{equation}
\hat{G}(\omega )=\sum_{{\bf k}j}\frac{\psi _{{\bf k}j\omega }^{l}\psi _{{\bf %
k}j\omega }^{r}}{\epsilon _{{\bf k}j\omega }-\omega }  \label{Gw}
\end{equation}
Note, that the Greens function is a non-hermitian matrix that is frequency
dependent so both the eigenvalues $\epsilon $ and right/left eigenvectors $%
\psi ^{r,l}$ are treated formally as frequency dependent quantities: $%
\epsilon _{{\bf k}j\omega },\psi _{{\bf k}j\omega }^{r}$, $\psi _{{\bf k}%
j\omega }^{l}.$ (The latter satisfies the Dyson equation (\ref{Dyson}) with
the wave function placed on the left.) In practice\cite{Anisimov,Nature}, (%
\ref{Dyson}) is solved on the Matsubara axis for a finite set of imaginary
frequencies $i\omega _{n}$ using some localized orbital representation such,
e.g., as linear muffin-tin orbitals (LMTOs) \cite{OKA}$\chi _{\alpha }^{{\bf %
k}}$ for the eigenvectors $\psi _{{\bf k}j\omega }^{r}$

\begin{equation}
\psi _{{\bf k}j\omega }^{r}=\sum_{\alpha }A_{\alpha }^{{\bf k}j\omega }\chi
_{\alpha }^{{\bf k}}  \label{Psi}
\end{equation}
which substitutes the differential equation (\ref{Dyson}) by a matrix
eigenvalue problem.

Once the local Green function is constructed, the new charge density,
effective potential and the local self-energy are computed. The latter is
found by solving the Anderson impurity model using a suitable many-body
technique. This entire formulation requires the self-consistent procedure
which delivers the total energy of the interacting electronic system.

The dynamical matrix is the second order derivative of the energy. As with
the ordinary density functional formulation of the problem\cite{SavrasovPRB}%
, we deal with the first order corrections to the charge density, $\delta
\rho ,$ as well as the first order correction to the local Greens function $%
\delta \hat{G}(\omega )$ which should be considered as two independent
variables in the functional of the dynamical matrix. To find the extremum, a
set of the linearized Dyson equations has to be solved self--consistently:

\begin{eqnarray}
(-\nabla ^{2}+V_{eff}+\hat{\Sigma}(\omega )-\hat{\Sigma}_{dc}-\epsilon _{%
{\bf k}j\omega })\delta \psi _{{\bf k}j\omega }^{r}+ &&  \nonumber \\
+(\delta V_{eff}+\delta \hat{\Sigma}(\omega )-\delta \hat{\Sigma}_{dc})\psi
_{{\bf k}j\omega }^{r}=0 &&  \label{Stern}
\end{eqnarray}
which leads us to consider the first order changes in the effective
potential $\delta V_{eff}$ and in the local self--energy operator $\delta 
\hat{\Sigma}(\omega ).$ Here and in the following we will assume that the
phonon wave vector of the perturbation ${\bf q}$ is different from zero,
and, therefore, the first order changes in the eigenvalues $\delta \epsilon
_{{\bf k}j\omega }$ drop out. The quantities $\delta V_{eff}$ and $\delta 
\hat{\Sigma}(\omega )$ are the functionals of $\delta \rho $ and $\delta 
\hat{G}(\omega )$ and should be found self--consistently. In particular, the
change in the self--energy $\delta \hat{\Sigma}(\omega )$ assumes the
development of solving an Anderson impurity model linearized with respect to
atomic displacements.

In practice, change in the eigenvector $\delta \psi _{{\bf k}j\omega }$ has
to be expanded in some basis set. Previous linear response schemes were
based on tight-binding methods \cite{Varma}, plane wave pseudopotentials 
\cite{BaroniPRL,Zein,GonzePRL,Quong}, linear augmented plane waves\cite
{KrakauerPRB}, mixed orbitals\cite{Bohnen} and linear muffin-tin orbitals 
\cite{SavrasovPRL}. To build an effective computational scheme applicable
for systems with localized orbitals we use LMTO representation as the basis.
Due to its explicit dependence on the atomic positions both Hellmann-Feynman
contributions and incomplete basis set corrections\cite{Pulay} appear in the
expression for the dynamical matrix \cite{SavrasovPRB}. We expand $\delta
\psi _{{\bf k}j\omega }$ as follows

\begin{equation}
\delta \psi _{{\bf k}j\omega }=\sum_{\alpha }\{\delta A_{\alpha }^{{\bf k}%
j\omega }\chi _{\alpha }^{{\bf k+q}}+A_{\alpha }^{{\bf k}j\omega }\delta
\chi _{\alpha }^{{\bf k}}\}  \label{dPsi}
\end{equation}
where we introduced both changes in the frequency dependent variational
coefficients $\delta A_{\alpha }^{{\bf k}j\omega }$ as well as changes in
the basis functions $\delta \chi _{\alpha }^{{\bf k}}.$ The latter helps us
to reach convergency in the entire expression (\ref{dPsi}) with respect to
the number of the basis functions \{$\alpha \}$ fast since the contribution
with $\delta \chi _{\alpha }^{{\bf k}}$ takes into account all rigid
movements of the localized orbitals \cite{SavrasovPRL}.

The first-order changes in the Green function can be found as follows 
\begin{equation}
\delta \hat{G}(\omega )=\sum_{{\bf k}j}\frac{\delta \psi _{{\bf k}j\omega
}^{l}\psi _{{\bf k}j\omega }^{r}+\psi _{{\bf k}j\omega }^{l}\delta \psi _{%
{\bf k}j\omega }^{r}}{\epsilon _{{\bf k}j\omega }-\omega }  \label{dGw}
\end{equation}
which should be used to evaluate the first order change in the charge
density and the dynamical matrix itself.

Despite formal complexity of the outlined method we see two significant
simplifications which can be implemented in practice. First, using static
limit of the dynamical mean field theory leads to a frequency independent
approximation for $\hat{\Sigma}.$ This recovers the LDA+U\ method and its
linear-response analog for the present problem. Second, it is generally
expected that only low frequency behavior of the self-energy will
drastically affect the static electronic response, since it modifies the
one--electron spectrum near the Fermi level. If the Hubbard subbands are
separated by sufficiently large $U$ they are essentially quasi-atomic
features and may be though as the states tightly bound and moving rigidly as
the atom vibrates. These two simplified calculational schemes may be
explored for the lattice dynamics in correlated materials without
significant complications.

We now describe our implementation of the method for calculating vibrational
spectrum of the prototype Mott insulator NiO. It is well known \cite{LDANiO}
that the LDA underestimates both the value of the energy gap and the
magnetic moment of this material. On the other hand, the use of the LDA+U
method fixes both problems \cite{LDANiO} when using non-zero Hubbard
parameter $U$. The use of the LDA+U\ approximation is equivalent here to a
static limit of the full frequency resolved theory outline above: the
self--energy $\hat{\Sigma}(\omega )$ becomes an orbital--dependent
correction to the Kohn-Sham potential $V_{eff}$ expressed via the density
matrix of the localized electrons, $n_{\alpha \beta }$, and the change in
the self--energy, $\delta \hat{\Sigma}$, is expressed via the changes in
occupation numbers, $\delta n_{\alpha \beta }$, due to atomic movements.
Here we will assume that the on-site Hubbard interaction $U$ is fixed for
all the displacements. The use of the static approximation significantly
simplifies our implementation: the solution of the impurity model and its
change with respect to the displacements collapses and the quantities which
have to be found self-consistently are the changes in charge density $\delta
\rho $ and in the occupancy matrix $\delta n_{\alpha \beta }.$ The latter
substitutes finding full change $\delta \hat{G}(\omega )$ in the local Green
function of the system.

We have calculated the phonon dispersions of NiO within the LDA and the
static limit of DMFT. We utilize 2$\kappa $ LMTO\ basis set for these linear
response calculations assuming antiferromagnetic spin alignment at Ni sites.
We use special point technique for integrating over the Brillouin zone
corresponding to a (6,6,6) grid and experimental lattice parameter a=7.926
a.u. LDA exchange--correlation after Ref. \cite{Vosko} \ is employed.

\begin{figure}[htb]
\epsfig{file=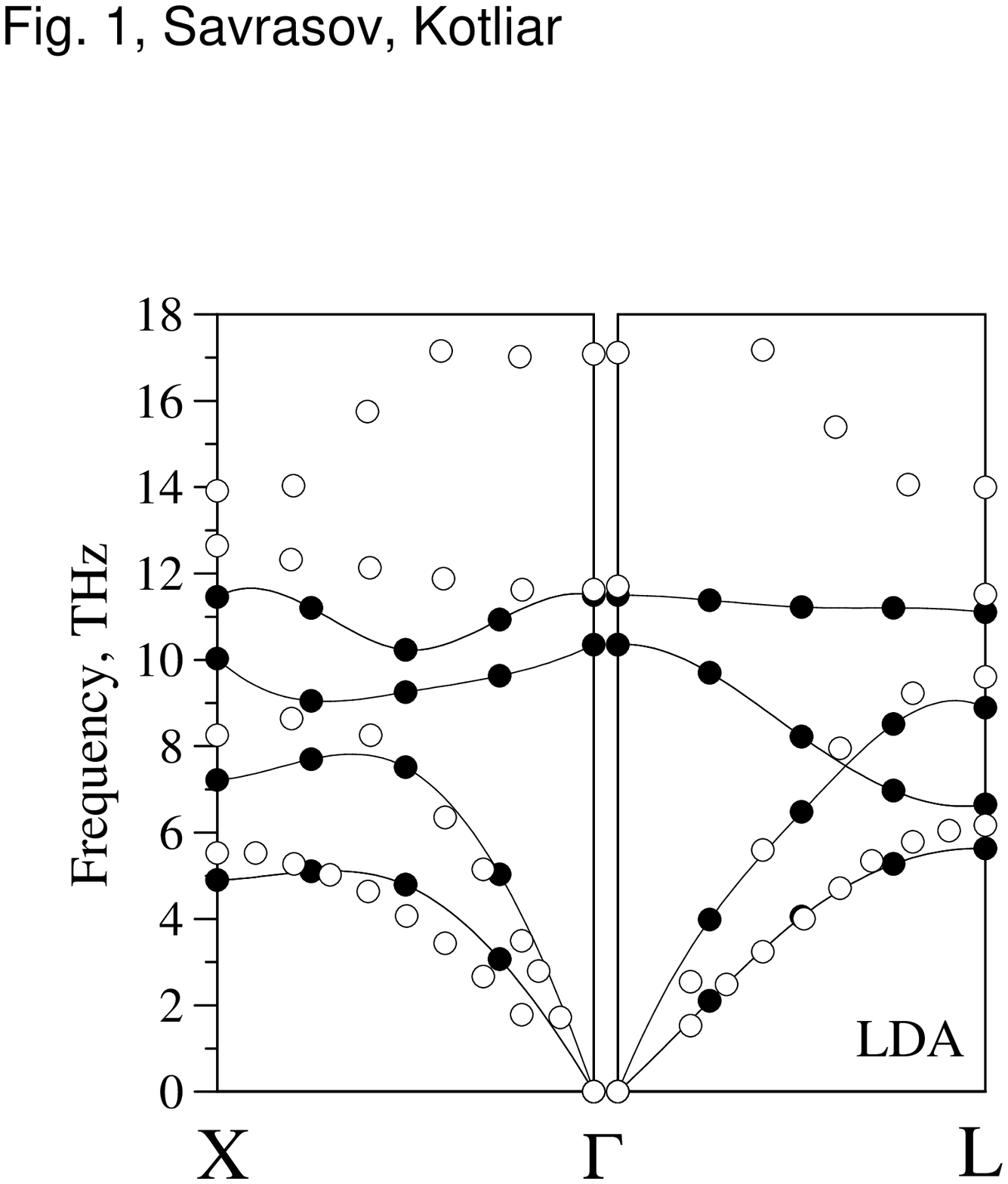,width=2.5in}
\caption{Comparison between calculated using the LDA (filled circles) and
experimental (open circles) phonon dispersion curves for NiO.}
\label{fig:One}
\end{figure}

\begin{figure}[htb]
\epsfig{file=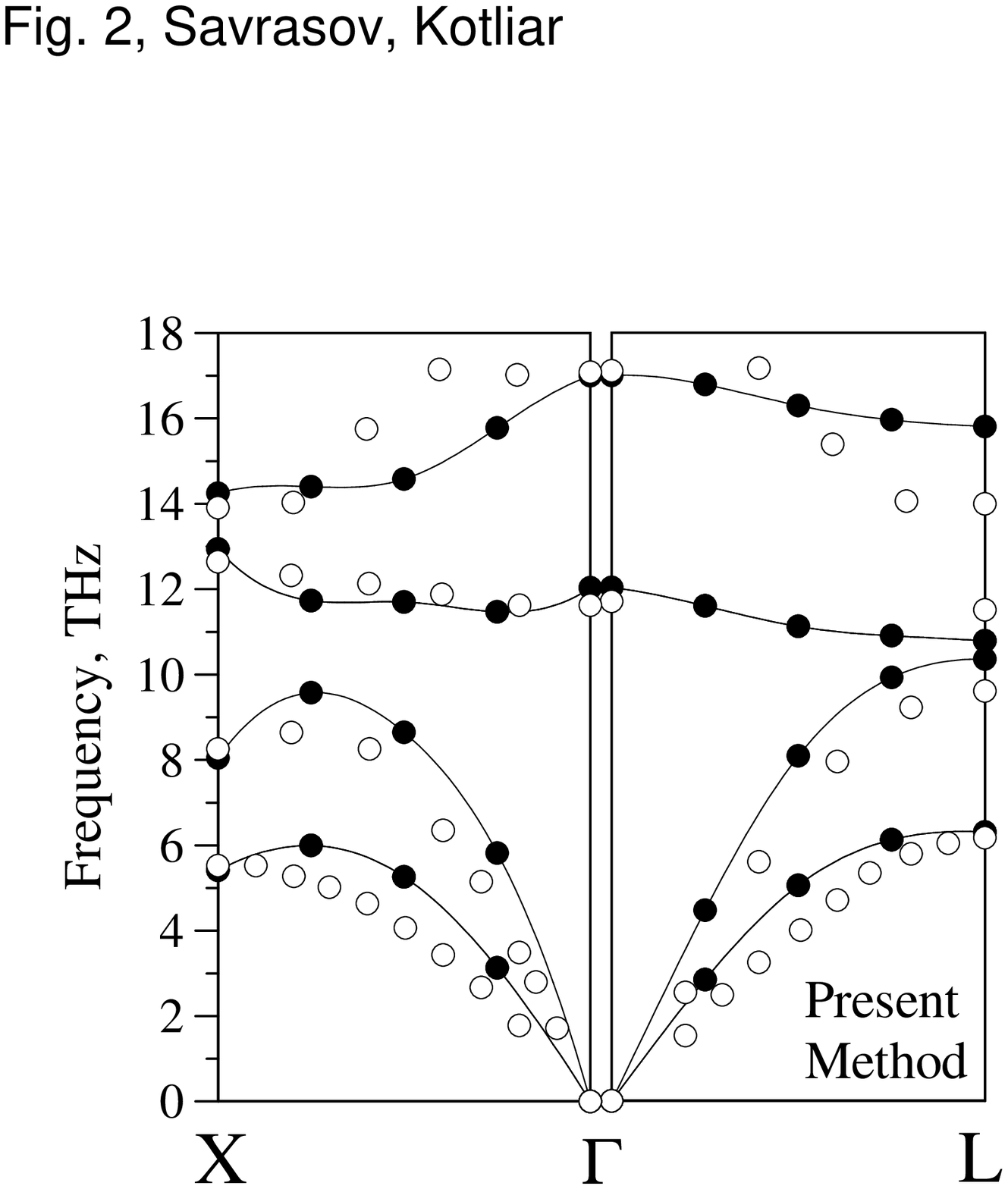,width=2.5in}
\caption{Comparison between calculated using the present method (filled
circles) and experimental (open circles) phonon dispersion curves for NiO.}
\label{fig:Two}
\end{figure}



We first discuss the results of the LDA calculations for the lattice
dynamics of NiO. Fig. 1 shows calculated phonon dispersion relations along
major symmetry directions. Despite the unit cell corresponding to
antiferromagnetic lattice contains two original unit cells, spin integrated
properties are unaffected by this symmetry lowering and therefore can be
represented in the original Brillouin zone. We see an apparent
underestimation of all theoretical optical modes by as much as 50 or more
per cent as compared to the experiment \cite{PhononNiO}. Similar finding for
MgO has also been reported \cite{Massida}. The LO--TO\ splitting is largely
underestimated which is predicted to be only 1.5 THz against 5 THz seen from
the measurement. Surprisingly, however, that both longitudinal and
transverse {\em acoustic} branches are much better predicted by the LDA. One
can conclude that the calculated sound velocities along $\Gamma X$ direction
are only slightly overestimated and the agreement is even better along the $%
\Gamma L$ line.

We now discuss how inclusion of correlations modifies the phonon spectrum.
Fig. 2 shows the calculated dispersion curves using static self--energy
approximation of the formalism developed in this work. We use the values of
Hubbard $U$=8 eV and exchange $J$=1 eV in this calculation. We see a
remarkable hardening of the optical modes which occurs due to proper
treatment of the correlational effects. Notably both the values of the
optical frequencies as well as the calculated LO--TO\ splitting are now in a
good agreement with the experiment. A pronounced softening of the
longitudinal optical mode along both $\Gamma X$ and $\Gamma L$ lines is seen
at the measured data which is in part captured by our theoretical
calculation: the agreement is somewhat better along the $\Gamma X$ direction
while the detailed q-dependence of these branches shows some residual
discrepancies.

Our acoustic modes are consistently overestimated by a few per cent as
compared to the experiment. They are also harden as the on-site interaction $%
U$\ increases although in much smaller degree than the optical modes.
Overall, the agreement between the theory and the experiment is much better
when the correlations are taken into account which indicates their
importance in the lattice dynamics of strongly correlated materials.

To better understand our findings we have calculated the values of $\epsilon
_{\infty }$ which represent an electronic contribution to the static
dielectric constant as given by the inverse element $1/\epsilon ^{-1}(0,0)$
of the full inverse dielectric permittivity matrix in reciprocal space$.$
Table 1 shows the results of these calculations using the LDA and our new\
method as well as the experiment \cite{EpsNiO}. It is seen that the LDA
result overestimates the value of $\epsilon _{\infty }$ by a factor of 6.
This is mainly because the calculated by LDA dielectric gap value is only
0.4 eV which is much smaller than the experimental energy gap of NiO equal
to 4 eV \cite{NiOgap}. Namely LDA overestimates the electronic screening
effects by a large amount causing both the artificial softening of optical
phonons and the lowering of the LO--TO splitting. The latter is well known
to be directly proportional to the Born effective charges but inversionally
proportional to $\epsilon _{\infty }.$

On the other hand, our\ calculations with correlations produce a much better
value of the static dielectric constant as seen from Table 1. We relate such
an agreement with the fact that the energy gap value predicted by the theory
here is 3.3 eV which compares well with the measurement \cite{NiOgap}. This
indicates that the screening effects in the calculation with correlations
are treated in much more appropriate way which fixes both the position of
the optical branches and their relative splitting.

As a final result, we have extracted the values of the Born effective charge 
$Z^{\ast }$ using ${\bf q}\rightarrow 0$ limit technique\cite{Krak} and the
relationship $\omega _{LO}^{2}-\omega _{TO}^{2}\propto |Z^{\ast
}|^{2}/\epsilon _{\infty }$. Table 1 shows comparison between the
theoretical\ and the experimental \cite{PhononNiO} data. The deviation from
2 which for binary oxides is the nominal rigid-ion value of $|Z^{\ast }|$
indicates high electronic polarizability. Both theoretical values are close
to the experimental one, while the agreement for LDA is apparently
accidental since the error in underestimating the splitting $\omega
_{LO}^{2}-\omega _{TO}^{2}$ nearly completely cancels the error in
overestimating $\epsilon _{\infty }.$

To summarize we have developed a new method to study the lattice dynamical
properties of materials with strong electronic correlations. The method is
based on the dynamical mean--field theory and it uses linear response
technique to calculate phonons at arbitrary wave vector ${\bf q}$. Using
static approximation we have computed the phonon dispersions of NiO and find
the results for the optical modes and static dielectric properties in much
better agreement with the experiment than the results of the corresponding
LDA based calculations. This emphasizes the importance of correlations in
lattice dynamics of strongly correlated systems. The residual discrepancies
can be attributed to the frequency dependence of the self--energy neglected
in the present study and will be a subject for the future work. There are
many other challenging problems for which the general framework outlined
here can be useful. Discrepancies between theory and experiment were noticed 
\cite{Vand} for the ferroelectric CaCu$_{3}$Ti$_{4}$O$_{12}$. The
electron-phonon interaction in cuprate superconductors is a subject of
intensive investigation. Finally we believe that correlation effects play an
important role in the phonon dynamics across the actinide series.

Acknowledgment: We are indebted to R. Resta and R. Cohen for valuable
discussions, and the Division of Basic Energy Sciences of the US\ DOE for
support under grant No. DE-FG02-99ER45761.

\begin{table}[tbp]
\caption{Comparison between calculated static dielectric constant $\protect%
\epsilon _{\infty }$ and Born effective charge $|Z^{\ast }|$ for NiO using
LDA and present method as well as the experimental data \protect\cite
{PhononNiO,EpsNiO} }
\label{I}
\begin{tabular}{cccc}
& LDA & Present Method & Exp \\ \hline
$\epsilon _{\infty }$ & 35.7 & 7.2 & 5.7, 6.1 \\ 
$|Z^{\ast }|$ & 2.17 & 2.33 & 2.22
\end{tabular}
\end{table}

\end{document}